\documentstyle[12pt]{article}
\renewcommand
\baselinestretch{1.4}

\def\beq{\begin{equation}}
\def\brr{\begin{array}}
\def\err{\end{array}}
\def\eeq{\end{equation}}
\def\bea{\begin{eqnarray}}
\def\eea{\end{eqnarray}}
\def\bs{\bigskip}
\def\tr{\mbox{Tr}\, }
\def\ni{\noindent}
\def\wt{\widetilde}
\def\wh{\widehat}

\def\nn{\nonumber}
\def\ms{\medskip}

\def\txs{\textstyle}
\def\dsp{\displaystyle}

\begin{document}

\hfill UB-ECM-PF 92/17
\mbox{}

\vspace*{1cm}

\begin{center}

{\LARGE \bf
One-loop divergences in two-dimensional Maxwell-dilaton quantum
gravity}

\vspace{8mm}

{\sc E. Elizalde and S.D. Odintsov}\footnote{On leave from
Department of
Mathematics and Physics,
 Pedagogical Institute, 634041 Tomsk, Russia.}
%\\ \mbox{} \\
\ms

Department E.C.M., Faculty of Physics, \\
University of Barcelona, \\
Diagonal 647, 08028 Barcelona, Spain \\
{\it e-mail: eli @ ebubecm1.bitnet}
\vspace{1cm}

{\sl June 1992}

\vspace{1cm}

{\bf Abstract}

\end{center}

Two-dimensional Maxwell-dilaton quantum gravity, which covers a large
family of the actions for two-dimensional gravity (in particular,
string-inspired models) is investigated. Charged black holes
which
appear in the theory are briefly discussed. The one-loop divergences in
the
linear covariant gauges are calculated. It is shown that for some
choices of the dilaton potential and dilaton-Maxwell coupling,
the
theory is one-loop multiplicatively renormalizable (or even finite). A
comparison with the  divergences structure of four-dimensional
Einstein-Maxwell gravity is given.

\vspace{1cm}

\noindent PACS: \begin{quote} 03.70 Theory of quantized fields, \
04.50 Unified theories and other theories of gravitation, \
11.10 Field theory, 11.17 Theories of strings and other extended
objects.
 \end{quote}

\newpage

\ni 1. \ {\sl Introduction}. As is known, the string-inspired
models of two-dimensional
gravity contain black hole solutions [1] as well as the Hawking
radiation [2], and can be exactly solvable classically [2]. These
theories (scalar-tensor or dilaton gravity) give us the
possibility
to have a very useful toy model which can tell us a lot about the
general properties of quantum gravity.

The study of the quantum properties of two-dimensional dilaton
gravity [3-10] shows that for some choices of  dilaton potential
the theory can be renormalizable [5-10], or even finite [8,10].
This can be compared with four-dimensional Einstein gravity,
which
is one-loop finite (on shell) but, unfortunately, is not
renormalizable. In the present letter we shall discuss quantum
corrections (divergences) in Maxwell-dilaton two-dimensional quantum
gravity with
the action
\beq
S=\int d^2x \, \sqrt{g} e^{-2\phi} \left[ R+ \gamma g^{\mu\nu}
\partial_{\mu} \phi \partial_{\nu}  \phi -\frac{1}{4} e^{\epsilon
(\phi) \phi} F_{\mu\nu}^2 + V( \phi ) \right], \label{s0}
\eeq
where  $F_{\mu\nu} =\partial_{\mu} A_{\nu}-\partial_{\nu}
A_{\mu}$ is the electromagnetic field-strength, $\phi$ the
dilaton, and $V(\phi)$ the dilaton potential. Particular cases
of action (\ref{s0}) describe a number of well-known models,
like the Jackiw-Teitelboim model [3], the bosonic string
effective action (for $\gamma =4$, $F_{\mu\nu} =0$, $V=\Lambda$)
or the heterotic string effective action (for $\epsilon =0$,
$V=\Lambda$) [11].

The model with the action (\ref{s0}) is connected with four or
higher-dimensional Einstein-Maxwell theories (or their
generalizations), which admit charged black hole solutions [12].
The theory (\ref{s0}) can also be considered as a toy model for
four-dimensional Einstein-Maxwell theory. Finally, as it has been
shown in refs. [11,13], for $\epsilon =$ const. and dilaton
potential of the type produced by string loops, the theory admits
charged black hole solutions with multiple horizons.

Actually, the study carried out with the action of [11] can
be generalized to the following one \cite{eo} (see also [13] for
constant $a$ and $b$)\footnote{In order to compare with the results in
[13], we use here the same Minkowski notations of [13] when discussing
the classical black hole solutions.}
 \beq
S=\int d^2x \, \sqrt{g} e^{-2\phi} \left[ R+ \gamma (\nabla \phi
)^2 -\frac{1}{4} a(\phi) F_{\mu\nu}^2 + b (\phi) (\nabla \psi )^2 + V(
\phi,
\psi ) \right], \label{s1}
\eeq
where
% $g = -\det g_{\mu\nu}$, $\psi $ is the (scalar) spectator
%field, $F_{\mu\nu} =\partial_{\mu} A_{\nu}-\partial_{\nu}
%A_{\mu}$ the electromagnetic field, and
$a(\phi)$ and $b(\phi)$ are arbitrary functions of the dilaton
field $\phi$.
The equations of motion for this generalized action are easily
obtained:
\bea
2 \nabla^2 \phi -4 (\nabla \phi )^2+ \frac{1}{4} a(\phi ) F_{\mu\nu}^2+
 V(\phi,\psi)&=&0, \nn \\
R_{\mu\nu}+2 \nabla_{\mu}\nabla_{\nu} \phi +(\gamma -4) \left[
\nabla_{\mu} \phi \nabla_{\nu} \phi- g_{\mu\nu} (\nabla \phi
)^2+\frac{1}{2} g_{\mu\nu} \nabla^2 \phi \right]
 - \frac{1}{2} a(\phi ) F_{\mu}^{ \ \lambda} F_{\nu\lambda}
\ \  & &
\\
+ \frac{1}{16} a^{(1)}(\phi ) g_{\mu\nu} F_{\alpha\beta}^2+ b(\phi )
\nabla_{\mu}
\psi\nabla_{\nu} \psi -\frac{1}{4} b^{(1)}(\phi ) g_{\mu\nu} (\nabla
\psi)^2 -\frac{1}{4}  g_{\mu\nu} \frac{\partial
V(\phi,\psi)}{\partial \phi} &=&0, \nn \\
\nabla^{\mu} \left[ F_{\mu\nu} a(\phi ) e^{-2\phi} \right] \, = \, 0,
\ \ \
2b(\phi ) \nabla^2 \psi + 2 [b^{(1)}(\phi )-2b(\phi )] \nabla \phi
\cdot
\nabla \psi - \frac{\partial V(\phi,\psi)}{\partial \psi}&=&0, \nn
\eea
where $a^{(1)}$ means derivative with respect to $\phi$.
In the case of static spherical configurations (also considered
in
\cite{ln}): \ $F=f(r)dr\wedge dt$, $\phi (r)$, $\psi (r)$, \,
with
an asymptotically flat metric \ $ds^2 =-g(r)dt^2+ g(r)^{-1}
dr^2$,
and \, $g(r)\rightarrow 1$ as $r \rightarrow \infty$,  we obtain:
\bea
(g\phi')'-2g(\phi')^2- \frac{a}{4} f^2+
\frac{1}{2} V (\phi,\psi) =0, \ \ \ \ \
2\phi'' + (\gamma -4) (\phi')^2+b(\psi')^2 &=& 0, \nn \\
\left( f a e^{-2\phi} \right)' =0, \ \ \ \ \
(g\psi')'+ \left( \frac{b^{(1)}}{b}-2\right) g\phi' \psi'-
\frac{1}{2b}
\, \frac{\partial V (\phi,\psi)}{\partial \psi} &=&0,
\eea
where the prime means derivative with respect to $r$.

However, in the present paper, the emphasis will be put in the
fact of allowing the Maxwell-dilaton coupling
 in (\ref{s1}) to be an arbitrary function of
the field $\phi$ (through the exponent $\epsilon (\phi) )$ and
---for the sake of simplicity of the discussion--- we shall not
take into account the dependence on the scalar spectator field
$\psi$.
% We shall now start with the discussion of the charged
%black hole solutions which generalized those of refs. [11,13] in
%the sense just described, for $\epsilon = \epsilon (\phi)$ an
%arbitrary function and for a different choice of $V(\phi)$, dictated by
%the renormalization structure.
 We shall now calculate the one-loop divergences in linear covariant
gauges and find the form of $V$ and $\epsilon (\phi)$ for which
the theory is one-loop renormalizable. Then, we use these $\epsilon
(\phi )$ and $V$ for the discussion of the charged black hole solutions
which generalize those of refs. [11,13].
 \bs

\ni 2. \ {\sl Calculation of the divergences}.
To this end, we write the action (\ref{s1}) (without $\psi$
field)
in a slightly different form, which is more convenient for the
investigation of quantum corrections:
\beq
S=\int d^2x \, \sqrt{g} e^{-2\phi} \left[ \frac{1}{2} g^{\mu\nu}
\partial_{\mu} \phi \partial_{\nu} \phi+ \gamma R -\frac{1}{4}
e^{\epsilon (\phi)\phi} F_{\mu\nu}^2  + V( \phi ) \right].
\label{s2}
\eeq
This action will be now our starting point. With the change of
field variable: \, $\wt{\phi}=e^{-\phi}$, \, we can write it  as follows
\beq
S=\int d^2x \, \sqrt{g}  \left[ \frac{1}{2} g^{\mu\nu}
\partial_{\mu} \wt{\phi} \partial_{\nu} \wt{\phi}+ \gamma
\wt{\phi}^2 R -\frac{1}{4} \wt{\phi}^{2-\epsilon (-\ln
\wt{\phi})}
F_{\mu\nu}^2  + \wt{\phi}^2 V(-\ln \wt{ \phi} ) \right].
\label{s3}
\eeq
Applying the transformation
\beq
c_1\varphi = \gamma \wt{\phi}^2, \ \ \ g_{\mu\nu} =
e^{2\rho}\wt{g}_{\mu\nu}, \ \ \ \wt{\phi} = \left( \frac{
c_1\varphi}{\gamma} \right)^{1/2}, \ \ \ \rho=\frac{\gamma
\wt{\phi}^2}{4c_1^2} - \frac{1}{8\gamma} \ln \wt{\phi},
\eeq
expression (\ref{s3}) can be rewritten in the way
\bea
S&=& \int d^2x \, \sqrt{\wt{g}} \left\{ \frac{1}{2}
\wt{g}^{\mu\nu}
\partial_{\mu} \varphi \partial_{\nu} \varphi+c_1
\varphi \wt{R}
+ e^{\varphi/2c_1} \left(
\frac{c_1\varphi}{\gamma}\right)^{1-
1/(8\gamma)} V \left( - \frac{1}{2} \ln\left(
\frac{c_1\varphi}{\gamma}\right)\right) \right.
\nn \\ &-& \left. \frac{1}{4} e^{-\varphi/2c_1} \exp
\left\{ \left[ 1-\frac{1}{2} \epsilon \left(- \frac{1}{2} \ln\left(
\frac{c_1\varphi}{\gamma}\right)\right)+ \frac{1}{8\gamma} \right]
\ln\left( \frac{c_1\varphi}{\gamma}\right) \right\} \wt{F}_{\mu\nu}^2
\right\},  \label{s4}
\eea
(where $\wt{F}_{\mu\nu}^2 = \wt{g}^{\mu\alpha} \wt{g}^{\nu\beta}
F_{\mu\nu} F_{\alpha\beta}$)
or, dropping the tilde,
\beq
S=\int d^2x \, \sqrt{g} \left[ \frac{1}{2} g^{\mu\nu}
\partial_{\mu} \varphi \partial_{\nu} \varphi+ c_1 \varphi R
-\frac{1}{4} f (\varphi) F_{\mu\nu}^2  + V( \varphi ) \right].
\label{s5}
\eeq
Here the first two terms represent the action of dilaton gravity,
the third one is the Maxwell term interacting with gravity, and
the
fourth one (the potential) has dimension $[V]= M^{-2}$. Note
that the action (\ref{s5}), with arbitrary $f$ and $V$, is the natural
generalization
of $d=4$ Einstein-Maxwell or higher derivative gravity-Maxwell theory.

The renormalization of the action (\ref{s5}) without the Maxwell
term has been investigated in linear covariant gauges in refs.
[5-7] and in the conformal gauge in [10]. It has been shown that the
theory
is renormalizable\footnote{Power counting shows that the theory is
renormalizable in the generalized sense (supposing the possible
change of $f$ and $V$ under renormalization).}
 for some choices of the
potential $V$ (in particular, for $V=\Lambda$
or $V=\mu e^{\alpha \varphi}$). Note that, generally speaking, there
exist gauges where the dilaton gravity is not one-loop
renormalizable, and others where it is a one-loop finite theory [8].

Our main purpose in this paper will be to calculate the one-loop
counterterms in the theory (Maxwell-dilaton gravity) given by the
action (\ref{s5}) in the linear covariant gauge (harmonic type
gauge).

We use the background field method. According to this procedure,
we
split the fields into their quantum and background parts:
\beq
g_{\mu\nu} \longrightarrow \bar{g}_{\mu\nu} =g_{\mu\nu}
+h_{\mu\nu}, \ \ \ \varphi \longrightarrow \bar{\varphi} = \phi +
\varphi, \ \ \ A_{\mu}  \longrightarrow \bar{A}_{\mu} =B_{\mu} +
A_{\mu},
\eeq
where the second terms $(h_{\mu\nu},\varphi,A_{\mu})$ are the
quantum fields.

Let us choose the gauge fixing actions in the gravitational and
electromagnetic field sectors, respectively, as the following
(linear gauge):
\bea
S^g_{GF} &=& -\frac{c_1}{2} \int d^2x \, \sqrt{g} \left(
\nabla_{\nu}h^{\nu}_{\mu}- \frac{1}{2} \nabla_{\mu} h -
\frac{1}{\phi} \nabla_{\mu} \varphi \right) \phi \left(
\nabla_{\rho}h^{\rho\mu}- \frac{1}{2} \nabla^{\mu} h -
\frac{1}{\phi} \nabla^{\mu} \varphi \right), \label{gf1} \\
S^{A_{\mu}}_{GF} &=& -\frac{1}{2} \int d^2x \, \sqrt{g} f(\phi )
\left( \nabla_{\mu} A^{\mu} \right)^2. \label{gf2}
\eea
One should add the gauge-fixing actions (\ref{gf1}) and
(\ref{gf2})
to the quadratic expansion ($S^{(2)}$) of (\ref{s5}) on the quantum
fields.

Now let us recall a  few simple expressions that are necessary
for the one-loop counterterms calculation. The one-loop
divergences of the euclidean effective action are given by
\beq
\Gamma_{div} = -\frac{1}{2}  \tr \ln \left. \wh{\cal
H}\right|_{div}+  \tr \ln \left. \wh{\cal M}_{gh}\right|_{div},
\label{gdi}
\eeq
where $\wh{\cal H}$ is defined through $S^{(2)} + S_{GF}$, and
$\wh{\cal M}_{gh}$ is the ghost operator corresponding to the
gauge fixing action.

If  $\wh{\cal H}$ has the following form
\beq
\wh{\cal H} = \wh{1} \Delta + 2 \wh{E}^{\lambda} \nabla_{\lambda}
+ \wh{\Pi}, \label{hpe}
\eeq
where  $\wh{\cal H}$ acts in the space of all quantum fields (as
well as $ \wh{1}$, $ \wh{E}^{\lambda}$ and $ \wh{\Pi}$), and
$\Delta= \nabla^{\mu}\nabla_{\mu}$; then, in dimensional
regularization,
\beq
  \tr \ln \left. \wh{\cal H}\right|_{div} = \frac{1}{\varepsilon}
\tr \left( \wh{\Pi}- \wh{E}^{\lambda}\wh{E}_{\lambda}\right).
\eeq
Here, the parameter of dimensional regularization is $\varepsilon =
2\pi (n-2)$. Notice that  we have dropped in (15) the
surface terms like $R$.
After the preceding remarks, we can now start the explicit calculation
of the one-loop counterterms.
\bs

\ni 2a. \ {\sl The dilaton-gravitational background (Maxwell
sector contributions)}.
In order to simplify the calculus we will do it in two steps.
In the first step we will be interested in the contribution of
the Maxwell sector to the action of dilaton gravity (the first
two terms and $V$ of (\ref{s0})). In this case we can put the
background vector field  $B_{\mu}=0$. Moreover, one can
immediately see that, due to the presence of only quadratic
terms in $A_{\mu}$, in $S^{(2)}$, the Maxwell sector decouples
from the gravitational sector. The corresponding term in the path
integral has the form (with account of the gauge fixing term
(12))
\beq
\int {\cal D} A^{\mu} \exp \left. \left( -\frac{1}{2} \int d^2x \,
\sqrt{g} f(\phi ) A_{\mu}
\wh{H}^{\mu\nu} A_{\nu} \right) \right|_{div} = \left. -
\frac{1}{2} \tr \ln \wh{H}^{\mu\nu} \right|_{div},
\eeq
where
\beq
 \wh{H}^{\mu\nu} = g^{\mu\nu} \Delta -  R^{\mu\nu}-
\frac{1}{f(\phi)} \left[ \delta_{\alpha}^{\mu} \left(
\nabla^{\nu} f(\phi) \right)- \delta_{\alpha}^{\nu} \left(
\nabla^{\mu} f(\phi) \right)- g^{\mu\nu} \left( \nabla_{\alpha}
f(\phi) \right) \right] \nabla^{\alpha},
\eeq
the local factor $f(\phi)$ not giving any contribution to
divergences. One can see that the operator  $ \wh{H}^{\mu\nu}$
has exactly the form (\ref{hpe}), with
\beq
\wh{\Pi} = -R^{\mu\nu}, \ \ \ \wh{E}_{\alpha} = -
\frac{1}{2f(\phi)} \left[  \delta_{\alpha}^{\mu} \left(
\nabla^{\nu} f(\phi) \right)- \delta_{\alpha}^{\nu} \left(
\nabla^{\mu} f(\phi) \right)- g^{\mu\nu} \left( \nabla_{\alpha}
f(\phi) \right) \right].
\eeq
The simple calculation of the one-loop counterterms with this $
\wh{H}^{\mu\nu}$ shows that there is no non-trivial contribution
from the Maxwell sector to the dilaton gravity sector (except for the
trivial surface terms which we consequently drop). The ghost
operator corresponding to the gauge fixing (12) is $\Delta$,
which again gives only  contribution  to the surface terms. Thus,
the complete Maxwell sector does not provide any contribution to
the quantum gravitational sector.
\bs

\ni 2b. \ {\sl Arbitrary vector and constant dilaton
background}.  We proceed now with the calculation of the one-loop
counterterm for the Maxwell action. For the sake of simplicity,
one can put $\phi=$ const. and  $g_{\mu\nu}= \delta_{\mu\nu}$ in
the background field splitting (10). Note that the
calculation of the one-loop counterterms in the gravitational
sector has been already done for the gauge under discussion in
[5,7]. It is not influenced by the addition of $F_{\mu\nu}$, as
we showed before; and that is why one can choose $g_{\mu\nu}=
\delta_{\mu\nu}$.

First of all, we should write the quadratic expansion of the
action (9) with account to the corresponding gauge fixing
terms. A straightforward (although lengthy) calculation gives, for
the background under consideration:
\beq
S^{(2)} + S^g_{GF} +S^{A_{\mu}}_{GF}= \frac{1}{2} \int d^2x \, \sqrt{g}
\ \left( A_{\rho} \ \
\bar{h}_{\mu\nu}  \ \ h \ \ \varphi \right) \wh{\cal H} \left( \brr{c}
A_{\tau} \\ \bar{h}_{\alpha\beta} \\ h \\ \varphi \err \right),
\eeq
the 16 components of $\wh{\cal H}$ being given by
\bea
H_{11} &=& f(\phi ) g^{\rho\tau} \Delta,
H_{12} = -f(\phi ) \left[ B^{\alpha\rho} \nabla^{\beta}-
\delta^{\rho\beta} B^{\alpha\sigma} \nabla_{\sigma} \right]-
\frac{1}{2} f(\phi ) \left[ (\nabla^{\beta}B^{\alpha\rho}) -
\delta^{\rho\beta} ( \nabla_{\sigma}B^{\alpha\sigma})  \right], \nn
\\
H_{13} &=& -\frac{1}{2} f(\phi )  B^{\sigma\rho} \nabla_{\sigma}-
\frac{1}{4} f(\phi )  (\nabla_{\sigma}B^{\sigma\rho}), \ \ \ \
H_{14} =  f^{(1)}(\phi )  B^{\sigma\rho} \nabla_{\sigma}+ \frac{1}{2}
f^{(1)}(\phi )  (\nabla_{\sigma}B^{\sigma\rho}), \nn \\
H_{21} &=& f(\phi ) \left[ B^{\nu\tau} \nabla^{\mu}-
\delta^{\tau\mu} B^{\nu\sigma} \nabla_{\sigma} \right]- \frac{1}{2}
f(\phi ) \left[ (\nabla^{\mu}B^{\nu\tau}) -\delta^{\tau\mu} (
\nabla_{\sigma}B^{\nu\sigma})  \right], \nn \\
H_{22} &=&  \frac{1}{2} \left( c_1 \phi \Delta - \wt{V} \right)
\wh{P}^{\mu\nu, \alpha\beta}- f(\phi) B^{\mu\sigma}B^{\rho}_{\ \sigma}
\delta^{\nu\tau} \wh{P}^{\alpha\beta}_{\rho\tau}, \ \ \ H_{23}
=-\frac{f(\phi)}{4} B^{\mu\sigma} B^{\nu}_{\ \sigma} = H_{32},
 \nn \\
H_{24} &=&  \frac{f^{(1)}(\phi)}{2} B^{\mu \sigma} B^{\nu}_{\
\sigma }=H_{42}, \    \
H_{31} = \frac{1}{2} f(\phi )  B^{\sigma\tau} \nabla_{\sigma}-
\frac{1}{4} f(\phi )  (\nabla_{\sigma}B^{\sigma\tau}), \ \  H_{33}=0,
\nn \\
 H_{34} &=&  \frac{1}{2} \left(-c_1 \Delta + \wt{V}^{(1)}
+\frac{f^{(1)}(\phi)}{2} B_{\mu\nu}^2 \right) =H_{43},
  \
H_{41} = -f^{(1)}(\phi )  B^{\sigma\tau} \nabla_{\sigma}+ \frac{1}{2}
f^{(1)}(\phi )  (\nabla_{\sigma}B^{\sigma\tau}), \nn \\
H_{44} &=&   \left(\frac{c_1}{\phi}-1\right) \Delta +
2\wt{V}^{(2)},
\eea
where $\wh{P} \equiv \delta^{\mu\nu,\alpha\beta} -\frac{1}{2}
g^{\mu\nu} g^{\alpha\beta}$, $h_{\mu\nu}\equiv \bar{h}_{\mu\nu}
+\frac{1}{2} \delta_{\mu\nu} h$, $B_{\mu\nu} \equiv \nabla_{\mu}
B_{\nu}-\nabla_{\nu} B_{\mu}$, $\wt{V} \equiv V(\phi ) -\frac{1}{4}
f(\phi ) B_{\mu\nu}^2$, and $ \wt{V}^{(1)}$ and $ \wt{V}^{(2)}$ are
the first and second derivatives of the potential with respect to
$\phi$, $\nabla_{\nu}$ being a flat derivative. The
contraction of the projector $\wh{P}$ of terms with $\alpha\beta$ and
$\mu\nu$ indices should also be done (because
$\bar{h}_{\alpha\beta} = \wh{P}_{\alpha\beta}^{\rho\tau}
\bar{h}_{\rho\tau}$,
$\bar{h}_{\mu\nu} = \wh{P}_{\mu\nu}^{\rho\tau}
\bar{h}_{\rho\tau}$).

One easily sees that the operator $\wh{\cal H}$ (20) does not have
the canonical structure (14) and, therefore, we cannot use the
algorithm (15). In order to render it possible the application
of this algorithm, we use a trick introduced in refs. [5,7]. We
express
\beq
\wh{H} = \wh{K} \wh{\cal H}, \label{red}
\eeq
where $\wh{K}$ is the constant matrix:
\beq
\wh{K} = \left( \brr{cccc} \displaystyle \frac{\delta^{\mu\nu}}{f(\phi
)} & 0 & 0 & 0 \\ 0 & \dsp \frac{2}{c_1\phi} \wh{P} & 0 & 0 \\
 0 & 0 & \dsp \frac{4}{c_1} \left( \dsp \frac{1}{c_1}- \dsp
\frac{1}{\phi}
\right) & \dsp - \frac{2}{c_1} \\ \displaystyle 0 & 0 & \dsp
-\frac{2}{c_1} & 0 \err \right), \eeq
and the operator $\wh{H} $ ---which can be easily evaluated from
(\ref{red})--- has now the canonical structure (14). It is
evident that, due to the fact that  $\wh{K}$ is a constant matrix, we
have \beq
\left. \tr \ln \wh{H} \right|_{div} = \left. \tr \ln \wh{\cal H}
\right|_{div},
\eeq
so that we can now concentrate on the operator $\wh{H}$ only.

Using the explicit form of  $\wh{H}$, we get
\beq
\wh{H} = \wh{1} \Delta + 2 \wh{E}^{\lambda} \nabla_{\lambda}
+ \wh{\Pi}, \label{hpe2}
\eeq
where the only non-zero components of the matrix $E_{\lambda}$ are:
\bea
(E_{\lambda})_{12} &=&  -\frac{1}{2} \left[ B^{\alpha \rho}
\delta^{\beta}_{\lambda}- B^{\alpha}_{ \ \lambda} \delta^{\rho\beta}
-\delta^{\alpha\beta} B_{\lambda}^{\ \rho}
\right], \ \ \
(E_{\lambda})_{13} =  -\frac{1}{4} B_{\lambda}^{ \ \rho}, \nn \\
(E_{\lambda})_{14} &=&   \frac{f^{(1)}(\phi )}{2 f(\phi )}
B_{\lambda}^{ \ \rho}, \  \
(E_{\lambda})_{21} =   \frac{f(\phi )}{c_1\phi } \left(
B^{\nu\tau} \delta^{\mu}_{\lambda}- B^{\nu}_{ \ \lambda}
\delta^{\tau\mu} -g^{\mu\nu} B^{ \ \tau}_{\lambda}  \right), \label{ela}
\\ (E_{\lambda})_{31} &=&   \frac{1}{c_1} \left[ \left(\frac{1}{c_1}-
\frac{1}{\phi} \right) f(\phi )+f^{(1)} (\phi ) \right]
B_{\lambda}^{ \ \tau}, \ \ \ \
(E_{\lambda})_{41} =  -\frac{f(\phi )}{2 c_1} B_{\lambda}^{ \ \tau}. \nn
\eea
Concerning the operator $\wh{\Pi}$, we need only its diagonal
components for the calculation of divergences. They are
\bea
\Pi_{11} =0, & & \Pi_{22} =
-\frac{\wt{V}}{c_1\phi} \wh{P}^{\alpha\beta,\mu\nu}
-\frac{2f(\phi)}{c_1\phi} \wh{P}^{\alpha\beta}_{\mu'\nu'}
B^{\mu'\sigma} B^{\rho}_{\ \sigma} \delta^{\nu'\tau}
\wh{P}^{\mu\nu}_{\rho\tau}, \nn \\
& & \Pi_{33} =\Pi_{44} =-\frac{\wt{V}^{(1)}}{c_1}
-\frac{f^{(1)}(\phi)}{2c_1} B^2_{\mu\nu}.
\eea
Now, it is straightforward to apply the algorithm (15) and
to calculate the one-loop divergences:
\beq
\Gamma_{div} = -\frac{1}{2} \tr \ln \wh{H} \label{div1}
= \frac{1}{\varepsilon} \int d^2x \, \sqrt{g} \left\{
\frac{V(\phi )}{c_1\phi} +\frac{V^{(1)}(\phi )}{c_1} -\frac{1}{2}
B_{\mu\nu}^2 \left[ \frac{f(\phi )}{2c_1^2} +\frac{f^{(1)} (\phi
)}{2c_1} \right]\right\}.
\eeq
Notice, as is easy to see, that the ghosts corresponding to the
gauge fixing actions (11) and (12) give  contributions to
the surface terms only. Here, the expression (\ref{div1}) is the
final one for the one-loop divergences of the effective action. We
can also see that the one-loop counterterms for the potential $V$
(which come from the gravitational sector)
coincide with the corresponding expressions found independently in
refs. [5,7]  (using the same gauge condition).

In order to write the final answer we should take into account the
counterterms of the gravitational sector (arbitrary
dilaton-graviton background). These counterterms have been
calculated in refs. [5,7]  (with the gauge fixing action (11), in
particular).
As it has been shown above (point 2a), the Maxwell sector does not
give here any contribution. So, finally, the one-loop divergences
of the effective action for the theory (9) are
\beq
\Gamma_{div} =  \frac{1}{\varepsilon} \int d^2x \, \sqrt{g}
\left[ \frac{3}{\phi^2} \left(\nabla_{\lambda} \phi \right)
\left(\nabla^{\lambda} \phi \right)
+\frac{V(\phi )}{c_1 \phi} + \frac{V^{(1)}(\phi )}{c_1} -
\frac{1}{4} B_{\mu\nu}^2 \left( \frac{f(\phi )}{c_1^2}
+\frac{f^{(1)}(\phi )}{c_1} \right)\right]. \label{div2}
\eeq
This expression constitutes the main result of our work.
\bs

\ni 2c. \ {\sl Renormalization}.   Let us now discuss the
renormalization of the theory under consideration. It follows from (28)
that the one-loop renormalized action is
\bea
S_R &=& \int d^2x \, \sqrt{g} \left[ \frac{1}{2} \left(1-
\frac{6}{\varepsilon \varphi^2} \right) g^{\mu\nu} \partial_{\mu}
\varphi \partial_{\nu} \varphi +c_1 \varphi R+ V(\varphi ) \left(1-
\frac{1}{\varepsilon c_1 \varphi} \right) \right. \nn \\
& -& \left. \frac{V^{(1)} (\varphi )}{\varepsilon c_1} -\frac{f(\varphi
)}{4} F_{\mu\nu}^2 \left( 1- \frac{1}{\varepsilon c_1^2} -
\frac{f^{(1)}(\varphi )}{\varepsilon c_1 f(\varphi )} \right)\right].
\label{ra1}
\eea
Choosing the one-loop renormalization of fields and constant in the
dilaton-gravity sector as the following (see [7] for a
discussion of this point)
\beq
\varphi   =\varphi_R, \ \ \ \ c_{1} = c_{1R}, \ \ \ \ g_{\mu\nu}
= e^{2\sigma (\varphi, c_1)} \wt{g}_{\mu\nu},
\eeq
where, in the one-loop approximation, $\sigma (\varphi, c_1)
=3/(2c_1\varepsilon \varphi )$, we obtain
\bea
S_R &=& \int d^2x \, \sqrt{\wt{g}} \left[ \frac{1}{2}   \wt{g}^{\mu\nu}
\partial_{\mu} \varphi \partial_{\nu} \varphi +c_1 \varphi \wt{R}+
V(\varphi ) \left(1- \frac{1}{\varepsilon c_1 \varphi} +
\frac{3}{\varepsilon c_1 \varphi} \right) \right. \nn \\
& -& \left. \frac{V^{(1)} (\varphi )}{\varepsilon c_1} -\frac{f(\varphi
)}{4} \wt{F}_{\mu\nu}^2 \left( 1- \frac{1}{\varepsilon c_1^2} -
\frac{3}{\varepsilon c_1 \varphi} -\frac{f^{(1)} (\varphi
)}{\varepsilon c_1 f(\varphi )} \right)\right], \label{ra2}
\eea
where $\wt{F}_{\mu\nu}^2 = \wt{g}^{\mu\alpha} \wt{g}^{\nu\beta}
F_{\mu\nu} F_{\alpha\beta}$, being $\wt{g}^{\mu\alpha}$ the
renormalized metric tensor.

It follows from  expression (\ref{ra2}) that, in the gauge under
consideration, Maxwell-dilaton gravity is one-loop renormalizable
for the following choices of the functions $V$ and $f$:
\bea
&& V (\varphi )= \mu \varphi^2 e^{-a \varphi}, \ \ \ \ \ \mu =
\left( 1-\frac{a}{\varepsilon c_1} \right) \mu_R, \nn \\
&& f (\varphi )= \frac{f_0}{\varphi^3} \exp \left[ -
\left( \frac{1}{c_1} +b \right) \varphi \right], \ \ \ \ \ f_0 =
\left( 1-\frac{b}{\varepsilon c_1} \right) f_0^R. \label{vfi}
\eea
Here $a$ and $b$ are arbitrary functions of $c_1$ (note that $\mu$
and $f_0$ can be choosen to be zero). Thus, we have shown that the
theory under consideration is one-loop multiplicatively
renormalizable in the gauge (11), (12), for special choices of $V$
and $f$ (\ref{vfi}).

By expanding in $e^{-\phi}$, this $V$ is
seen to correspond to a special case of the type of dilaton potential
one expects from closed string loop corrections [11]. In fact, in
terms of the original dilaton field $\phi$ (see (5)), we get
\beq
V(\phi )=  \mu' \, e^{\txs \alpha \phi} \, e^{\txs A'e^{\txs -2\phi}}, \
\ \ \ \
\ \ \ A(\PHI )=  A' \, E^{\TXS -(\ALPHA-6) \PHI}\ E^{ \TXS
B'e^{\txs -2\phi}}. \eeq
These are indeed very nice functions of the variable $\phi$. In
particular, for
positive values of all the coefficients involved, $V$ develops a minimum
for a rather small value of $\phi$ (depending, of course, on the precise
values of the constants), while $a$ has a monotonical, exponentially
decreasing shape. When $A'$ and $B'$ are negative, $V$ is an
exponentially increasing function of $\phi$ while $a$ has a maximum
previous to an exponential decrease. $V$ also develops a minimum for
negative values of $\mu'$, $\alpha$ and $A'$. Notice that if one uses
the expansion on $\phi$ of $\exp (\exp (-2\phi))$ (as in the
perturbative case of refs. [11,13]), then the minimum is never seen.

Let us now compare the one-loop structure of the theory (9)
with the structure of four-dimensional Maxwell-Einstein theory,
which is known to be one-loop non-re\-nor\-mal\-iz\-able [15], even on
shell. First of all, one can see that there is the contribution to
the gravitational sector ($R^2$-terms) from the electromagnetic
sector in four dimensions [15]. As we have seen, in the
two-dimensional case there is no contribution to the
dilaton-gravity sector. Second, while there is no contribution from
the gravitational sector to $F_{\mu\nu}^2$ in four dimensions,
here, on the contrary, we see that for Maxwell-dilaton gravity
there is such contribution from the dilaton-gravity sector, even if
the function $f(\varphi )$ is choosen to be constant. Third, new
counterterms appear [15] which differ from the original action and
which, in the end, lead to the non-renormalizability of
four-dimensional Maxwell-Einstein theory even on shell. Such terms
have not appeared in the theory (9), which is one-loop
multiplicatively renormalizable off-shell for $V$ and $f$ given in
(\ref{vfi}). (Note that if, moreover, we take $a=0$ and $b=0$ in (33),
then two-dimensional Maxwell-dilaton gravity is actually one-loop
finite,\footnote{Perturbative properties (in particular, two-loop
finiteness) of the supersymmetric extension of matter-dilaton gravity
($d=2$ supergravity) have been recently investigated in ref. [19].}
 compare with [8,10]). Finally, we must remark that for the
two-dimensional theory under consideration, as well  as for the
four-dimensional Maxwell-Einstein theory, one-loop counterterms
(off-shell) depend on the gauge. Here, presumably, the use of a
gauge independent effective action (see [16,17] for a discussion of
gauge independent effective action in two dimensional gravity) can
add some information on the renormalization structure of the
theory.
\bs

\ni 2d. \ {\sl Charged black holes}.  The solutions of the two
last eqs. (4) (when $\psi$ is absent from the begining) are:
\beq
f(r)=f_0 a^{-1} (\phi (r)) e^{2\phi (r)}, \ \ \ \
\phi (r )= \left\{
\brr{ll}
\phi_0 + \frac{\alpha}{2} \ln r,  &  \mbox{for} \ \gamma\neq 4,
\\ \phi_0 - \frac{Q}{2} \, r,   & \mbox{for} \ \gamma = 4, \err
\right. \eeq
where $\alpha \equiv 4/(\gamma -4)$, and
$f_0$, $\phi_0$ and $Q$ are arbitrary constants (see also [13]). The
solution for $g(r)$ (coming from the first of eqs. (4)) is
\beq
g(r)= \left\{
\brr{ll}
r^{\alpha +1} \left[ -2m- \frac{1}{\alpha} \int^r ds \, s^{-\alpha}
W(\phi (s)) \right], & \mbox{for} \ \gamma\neq 4, \\
e^{-Qr} \left[ -2m+ \frac{1}{Q} \int^r ds \, s^{-\alpha}
W(\phi (s)) \right], & \mbox{for} \ \gamma = 4, \err \right.
\eeq
where $W(\phi )\equiv V(\phi ) -f_0^2e^{4\phi}/(2a(\phi))$ and $m$ is
a new constant (it turns out to be the black hole's mass).

This solution for the potential (33) describes a Reissner-Nordstr{\o}m
black hole.
We do not discuss the properties of (34)-(35) here due to lack of space.
\bs

In summary, we have studied the renormalization structure of
two-dimensional Maxwell-dilaton quantum gravity and showed that
this theory can be one-loop multiplicatively renormalized. This
theory admits also charged black holes. However, it seems [13,18] that
these solutions do not satisfy scalar no-hair theorems. In order to
have these theorems we ought to introduce scalar spectator fields
[13] (as in eqs. (2)), which arise
in the superstring context  from the Ramond-Ramond
field. Then it would be interesting to investigate the one-loop
structure of such theory. Work along this line is in progress.

\vspace{5mm}

\ni{\large \bf Acknowledgments}

We would like to thank Profs. Ignatios Antoniadis, Ron Kantowski and
Andrei Slavnov for fruitful discussions on connected problems, and Ali
Chamseddine and Arkadii Tseytlin for correspondence.
S.D.O. thanks the members of the Department E.C.M. of
Barcelona University for hospitality and
E.E. the Alexander von Humboldt Foundation for continued help.
This work has
been supported by Direcci"n Ge\-ne\-ral de Investigaci"n
Cient!fica y Tcnica (DGICYT), research projects
 PB90-0022 and SAB92-0072.

\newpage

\renewcommand
\baselinestretch{1.05}

\end{document}